# Dynamic and Auto Responsive Solution for Distributed Denial-of-Service Attacks Detection in ISP Network

B. B. Gupta, *Student Member, IEEE,* R. C. Joshi, and Manoj Misra, *Member, IEEE*

***Abstract*—Denial of service (DoS) attacks and more particularly the distributed ones (DDoS) are one of the latest threat and pose a grave danger to users, organizations and infrastructures of the Internet. Several schemes have been proposed on how to detect some of these attacks, but they suffer from a range of problems, some of them being impractical and others not being effective against these attacks. This paper reports the design principles and evaluation results of our proposed framework that autonomously detects and accurately characterizes a wide range of flooding DDoS attacks in ISP network. Attacks are detected by the constant monitoring of propagation of abrupt traffic changes inside ISP network. For this, a newly designed flow-volume based approach (FVBA) is used to construct profile of the traffic normally seen in the network, and identify anomalies whenever traffic goes out of profile. Consideration of varying tolerance factors make proposed detection system scalable to the varying network conditions and attack loads in real time. Six-sigma method is used to identify threshold values accurately for malicious flows characterization. FVBA has been extensively evaluated in a controlled test-bed environment. Detection thresholds and efficiency is justified using receiver operating characteristics (ROC) curve. For validation, KDD 99, a publicly available benchmark dataset is used. The results show that our proposed system gives a drastic improvement in terms of detection and false alarm rate.**

## I. INTRODUCTION

Nowadays, DDoS attack is the most disastrous and difficult threat to the Internet. It is commonly characterized as an event in which a large number of unwitting hosts are used as an attack force against the victim to exhaust either their computational or communication resources. As a result, legitimate users are denied from the services that they normally expect to have [1]. Therefore, as given by Weiler [2] DDoS attacks attempt: (1) to inhibit legitimate network traffic by flooding the network with useless traffic. (2) to deny access to a service by disrupting connections between two parties. (3) to block the access of a particular individual to a service. (4) to disrupt the specific system or service itself. Intruder can perform DDoS attacks either as flooding attacks or as logical attacks. These attacks reveal big loopholes not only in specific applications, but also in the entire TCP/IP protocol suite. Series of DDoS attacks that shut down some high profile websites have demonstrated the severe consequences of these attacks [3]. As per computer crime and security survey conducted by FBI/CSI in the United States for the year 2004 [4], these attacks are the second most widely detected outsider attack types in computer networks immediately after virus infections. A computer crime and security survey conducted in Australia for the year 2004 [5] shows similar results. Technologies, such as cable modems to home users, have further increased the risk of DDoS attacks. This is because with the cable modems the home users are always connected to the Internet and it is easier for an attacker to compromise these systems, which often have weak security [6].

Currently, the majority (90-94%) of DDoS attacks are performed using TCP, and a large portion (52-57%) of them is targeted to flooding attacks [7]. Therefore, we concentrate on thwarting a wide range of flooding attacks. Flooding DDoS attacks performed by attackers can be broadly categorized in following three categories:

High rate disruptive: In high rate disruptive attacks, sheer volume of packets at very high rate are sent from distributed locations in a coordinated manner to completely disrupt the availability of Internet services. As these attacks have direct impact on ISP networks, thus easy to detect and characterize.

Diluted low rate degrading: In diluted low rate degrading attacks, packets are sent from a large number of infected machines i.e. zombie machines at low rate in a coordinated manner to gracefully degrade network performance. As these attacks degrade Quality of Service (QoS) of the network slowly, thus very difficult to detect and characterize.

Varied rate: To make detection of attacks more difficult, attackers can use some sophisticated attack tools to generate varied rate attacks in which they use some of the zombie machines to generate packets at high rate while remaining machines to generate packets at low rate. This type of attacks is toughest to detect and characterize from entropy based approaches.

Over the recent years, several schemes have been proposed to detect flooding DDoS attacks [8]-[22]. Most of them are able to detect high rate disruptive attacks while ineffective

Manuscript received March 1, 2009. This work was supported in part by the Ministry of Human Resource Development, Government of India.

B. B. Gupta is with the Department of Electronics & Computer Engg., Indian Institute of Technology, Roorkee, 247667 India. (phone: +91-9927713132; e-mail: bbgupta@ieee.org).

R. C. Joshi is with the Department of Electronics & Computer Engg., Indian Institute of Technology, Roorkee, 247667 India.

Manoj Misra is with the Department of Electronics & Computer Engg., Indian Institute of Technology, Roorkee, 247667 India.

against other flooding attacks. This is because of difficulty in constructing profile of normal traffic to detect anomalies, when system is under diluted low rate degrading attacks. Entropy based approaches [19]-[21] can detect diluted low rate degrading attacks, but failed against varied rate attacks wherein intelligent attacker mixes low and high rate zombie machines to generate attack traffic in such a manner that overall entropy remain unchanged.

In this paper, we present a novel framework that concentrates on detection and characterization of a wide range of flooding DDoS attacks, e.g. high rate disruptive, diluted low rate degrading and varied rate, by monitoring the propagation of abrupt traffic changes inside ISP network. Early version of our framework is presented in [23]. A newly designed Flow-Volume Based Approach (FVBA) is used to construct profile of the traffic normally seen in the network, and identify anomalies whenever traffic goes out of profile. Proposed detection system is scalable to varying network conditions and adapts itself to different attacks loads. Six-sigma [24], [25] method is used to identify threshold values accurately for malicious flows characterization. Internet type topologies used for simulation are generated using Transit-Stub model of GT-ITM [26] topology generator. NS-2 [27] network simulator on Linux platform is used as simulation test-bed. A publicly available benchmark dataset, KDD 99 [28], is used to validate the efficiency and effectiveness of proposed approach. The results show that our proposed scheme inflicts an extremely high detection rate with low false alarm rate.

The remainder of the paper is organized as follows. Section II briefly reviews related work. Section III describes proposed approach and its advantages in the details. Section IV contains experimental setup and performance analysis. Validation of proposed approach with real data is reported in section V. Finally, Section VI concludes the paper and outlines future work.

## II. Related Work

This section charts out the overview on a plethora of existing DDoS defense schemes proposed in the literature.

Exiting DDoS defense mechanisms are classified into four broad categories: Prevention, Detection, Response, and Tolerance & mitigation. However, research on DDoS attacks is primarily focused on attack detection and response mechanisms. Attack detection aims to detect an ongoing attack and to discriminate malicious traffic from legitimate traffic. Typical detection techniques fall into four categories: signature based attack detection, anomaly based attack detection, hybrid attack detection and third party attack detection. In the signature based detection techniques, database of attack signatures is used to detect attacks. The signatures are manually constructed by security experts analyzing previous attacks and used to match with incoming traffic to detect intrusions. SNORT [8] and Bro [9] are the two widely used signature based detection approaches. Signature based techniques are only effective in detecting traffic of known DDoS attacks whereas new attacks or even slight variations of old attacks go unnoticed.

Anomaly based detection techniques on the other hand, relies on constructing profile of valid traffic patterns and identifies anomalies whenever traffic goes out of profile. Most of DoS detection systems [10]-[20] are anomaly based. In [10], Gil and Poletto proposed a scheme called MULTOPS to detect denial of service attacks by monitoring the packet rate in both the up and down links. MULTOPS assumes that packet rates between two hosts are proportional during normal operation. A significant disproportion between the packet rate going to and from a host or subnet is a strong indication of a DoS attack. Blazek et al. [11] proposed batch detection to detect DoS attacks by monitoring statistical changes. Lee and Stolfo [12] used data mining techniques to discover patterns of system features that describe program and user behavior and implement a classifier that can recognize anomalies and intrusions. A mechanism called congestion triggered packet sampling and filtering is proposed by Huang et al. [13]. According to this approach, a subset of dropped packets due to congestion is selected for statistical analysis. If anomaly is indicated by the statistical results, a signal is sent to the router to filter the malicious packets. Cheng et al. [14] proposed to use spectral analysis to identify DoS attack flows. Mirkovic et al. [15] proposed D-WARD defense system that does DDoS attack detection at source, based on the idea that DDoS attacks should be stopped as close to the source as possible. In [16], flow belonging to DoS attacks is identified by considering high traffic volume to the victim. Then right drop probability for such traffic is calculated and conveys this information to the upstream routers, which in turn could drop packets belonging to the attack traffic themselves. Bencsath et al. [17] have given a traffic level measurement based approach, in which incoming traffic is monitored continuously and dangerous traffic intensity rises are detected. Chen et al. [18] used distributed change-point detection (DCD) architecture using change aggregation trees (CAT) to detect DDoS attack over multiple network domains. Anomaly based techniques can detect novel attacks; however, it may result in higher false alarms.

Hybrid detection approaches [22] combine the positive features of the signature and anomaly based detection models to achieve high detection rate with low false positives. Even though it decreases false positive rate but complexity and cost of implementation of hybrid attack detection system is very high. Mechanisms that deploy third-party detection do not handle the detection process themselves, but rely on an external third-party that signals the occurrence of the attack [23]. Examples of mechanisms that use third-party detection are easily found among traceback mechanisms [29]-[31]. Though most of DoS detection systems [8]-[18], use volume based metrics to detect and characterize DDoS attacks and have been successful in countering high rate disruptive flooding attacks, but diluted low rate flooding attacks can not be detected and characterized because these attacks do not cause detectable disruptions in traffic volume. These suffer in the form large number of false positives and false negatives hence more collateral damage when attack is carried at slow rate or when volume per attack flow is not so high as compared to legitimate flows. On the other hand, entropy based DDoS detection approaches in [19]-[21] can counter diluted low rate degrading flooding attacks too, but ineffective against varied rate attacks.

Attack response systems attempt to alleviate the damage caused by the attacks by taking proactive measures [32], [33] or reactive measures like localizing the source of the attacks using traceback approaches [29]-[31], by filtering malicious traffic if characterized correctly [34], or reducing intensity of attacks using rate throttling approaches [15], [16], [35].

## III. Our Approach

After analyzing various existing DDoS defense techniques, we find that major challenge of defense against DDoS attacks is how to detect and identify the attack traffic accurately and efficiently. Therefore, by considering this challenge proposed framework aims to provide following activities in thwarting DDoS attacks:

**Detection:** Detects a wide range of flooding DDoS attacks i.e. high rate disruptive, diluted low rate degrading and varied rate autonomously in ISP network while victim is being attacked

**Malicious Flows Characterization:** Identifies and tags attack flows accurately in real time.

Rest of the section describes proposed framework in details.

### A. Detection

Here, we will discuss propose detection system that is part of access router or can belong to separate unit that interact with access router to detect attack traffic. Detecting DDoS attacks involve first knowing normal profile of the system and then to find deviations from this normal profile. Whenever incoming traffic goes out of the normal profile, anomalous system behavior is identified. Our approach detects flooding DDoS attacks by the constant monitoring the propagation of abrupt traffic changes inside the ISP network. Various statistical measures e.g. volume, flow, entropy, ratio etc are available for profile generation. A high-level block diagram of DDoS detection system is given in fig. 1.

A newly designed flow-volume based approach (FVBA) is used to construct profile of the traffic normally seen in the network, and identify anomalies whenever traffic goes out of profile. In FVBA, two statistical measures namely volume and flow are used for profile construction. Fig. 2 depicts the FVBA architecture, where $X_{in}(t)$ represents total traffic arriving at the target machine in Δ time duration, when system is under attacks. $X_{in}(t)$ can be expressed as follows:

$$X_{in}(t) = X_n^*(t) + \hat{\hat{X}}(t), \tag{1}$$

where, $X_n^*(t)$ and $\hat{\hat{X}}(t)$ are the components of the normal and attack traffic respectively.

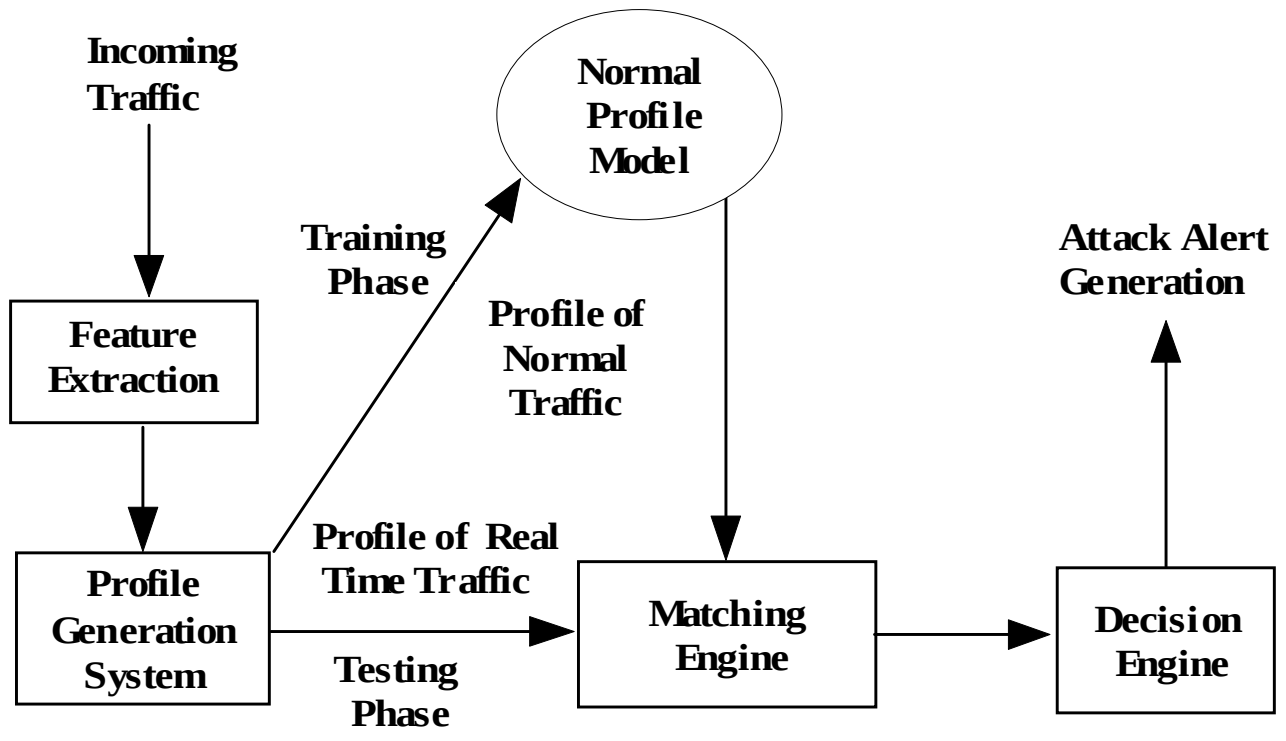

Fig. 1 High-level block diagram of DDoS detection system

$X_{in}(t)$ - $X_n^*(t)$ using above equation can be used for detection purpose.

To set normal profile, consider a random process { $X(t)$ , $t = w\Delta$ , $w \in N$ }, where Δ is a constant time interval, $N$ is the set of positive integers, and for each $t$, X ($t$) is a random variable. $1 \le w \le l$ , $l$ is the number of time intervals. Here $X$ ($t$) represents the total traffic volume in {$t$ − Δ, $t$} time interval. $X$ ($t$) is calculated during time interval {$t$ − Δ, $t$} as follows:

$$X(t) = \sum_{i=1}^{Nf} n_i, i = 1, 2....N_f \tag{2}$$

where $n_i$ represent total number of bytes arrivals for a flow i in {$t$ − Δ, $t$} time duration and $N_f$ represent total number of flows. We take average of $X$ ($t$) and designate that as $X_n^*(t)$ normal traffic Volume. Similarly value of flow metric is calculated and designates that as $F_n^*(t)$. Here total bytes, not packets, are used to calculate volume metric, because it provides more accuracy, as different flows can contain packets of different sizes.

To detect the attack, the value of volume metric $X_{in}(t)$ and flow metric $F_{in}(t)$ is calculated in time window Δ continuously; whenever there is appreciable deviation from $X_n^*(t)$ and $F_n^*(t)$, various types of attacks are detected using algorithm 1 as given in fig. 3. Threshold values $x_{th}$, $x_{th}^L$ and $V_{th}$ are set as follows:

$$x_{th} = r_1 * s_V \tag{3}$$

$$V_{th} = r_2 * s_F \tag{4}$$

$$x_{th}^L = r_3 * s_V \tag{5}$$

where $s_V$, $s_F$ represents value of standard deviation for volume measure and flow measure, respectively. $r_1, r_3 \in I$ , represent value of upper and lower bound of tolerance factor for volume measure, respectively, where $I$ is a set of integers. $r_2 \in I$ , represent value of tolerance factor for flow measure. Effectiveness of an anomaly based detection system highly depends on accuracy of threshold value settings. Inaccurate threshold values cause a large number of false positives and false negatives. Therefore, various simulations are performed using different value of tolerance factors. The choice of tolerance factors varies for different network conditions. Values of tolerance factors also depend on the composition of the normal traffic and the desired degree of the ability to control a DDoS attack.

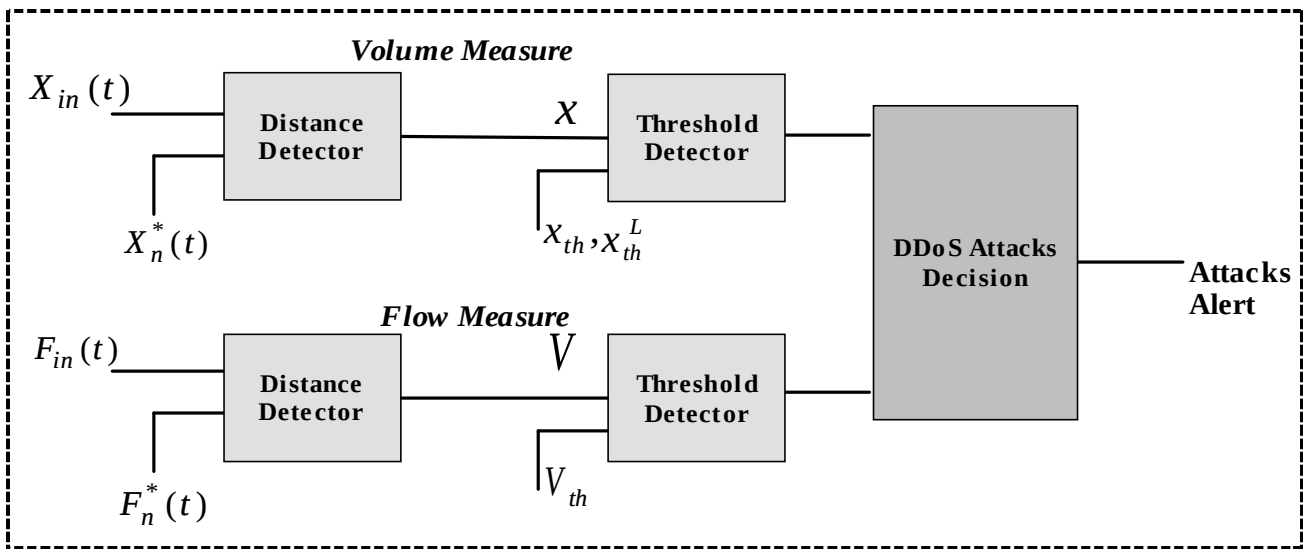

Fig. 2 FVBA architecture

---

***Algorithm 1: DDoS attacks Detection Algorithm***

***Input:*** *$X_n^*(t)$ : Normal traffic Volume measure*

*$F_n^*(t)$ : Normal traffic Flow measure*

*$x_{th}$ : Upper bound of Threshold value for Volume measure*

*$V_{th}$ : Threshold value for Flow measure*

*$x_{th}^L$ : Lower bound of threshold value for Volume measure*

***Output:*** *DDoS attack alert generation.*

***Procedure:***

***01:*** *Consider a random process { $X_{in}(t)$ , $F_{in}(t)$ , $t = w\,\Delta$ , $w \in N$ }, where Δ is a constant time interval, N is the set of positive integers, and for each t , $X_{in}(t)$ and $F_{in}(t)$ are random variables. $1 \le w \le l$ , $l$ is the number of time intervals. Here, $X_{in}(t)$ represents value of volume measure and $F_{in}(t)$ represents value of flow measure in time duration {t−Δ, t}.*

***02: If*** *(( $X_{in}(t) - X_n^*(t) > x_{th}$ ) || ( $F_{in}(t) - F_n^*(t) > V_{th}$ ))* ***Then*** *Attack pattern detected. // in case of TCP and ICMP flooding attacks*

*DDoS attack alert is generated.*

***03: Else If*** *(( $X_{in}(t) - X_n^*(t) < x_{th}$ ) && ( $F_{in}(t) - F_n^*(t) < V_{th}$ ))* ***Then*** *System is attack free.*

*Attack alert is not generated.*

***04: If*** *(( $X_{in}(t) - X_n^*(t) > x_{th}$ ) || ( $X_{in}(t) - X_n^*(t) < x_{th}^L$ ) || ( $F_{in}(t) - F_n^*(t) > V_{th}$ ))* ***Then*** *Attack pattern detected. // in case of UDP flooding attack*

*DDoS attack alert is generated.*

***05: Else If*** *(( $X_{in}(t) - X_n^*(t) < x_{th}$ ) && ( $X_{in}(t) - X_n^*(t) > x_{th}^L$ ) && ( $F_{in}(t) - F_n^*(t) < V_{th}$ ))* ***Then*** *System is attack free.*

*Attack alert is not generated.*

---

Fig. 3 Algorithm for DDoS attacks Detection

Usually UDP traffic is conducted using few connections and occupies a small percentage of all the network traffic therefore lower bound of volume measure is also used for UDP type of attacks detection.

### *B. Malicious Flows Characterization*

Our aim in this paper is to detect and characterize a wide range of DDoS attacks in ISP network under varying network conditions and validate proposed system with real dataset. For the sake of completion, we describe the characterization mechanism, as proposed in [23], in this section.

Once attacks occurrence is detected, next thing to do is correctly separation of traffic coming through malicious flows from legitimate traffic to respond to attacks correctly. For this, we observed total number of the bytes arrival for each flow during monitoring period, and flows that crosses predefined thresholds are classify either suspicious or attack traffic flows depending on deflection from thresholds. Assume $F$ is the set of active flows, then ( $F = F_{normal} \cup F_{attack}$ ) AND $(F_{normal} \cap F_{attack} = f)$ , where $F_{normal}$ represent set of actual normal flows and $F_{attack}$ is set of actual attack flows.

Characterization algorithm outputs subsets $F_{attack}^*$ , $F_{suspicious}^*$ of $F$ . Here, $F_{attack}^*$ and $F_{suspicious}^*$ represent set of attack and malicious flows respectively, given as output by our characterization algorithm. Ideally $(F_{attack}^* \cap F_{attack} = F_{attack}) AND (F_{attack}^* \cap F_{normal} = f)$ and $(F_{suspicious}^* \cap F_{normal} = f) AND (F_{suspicious}^* = f)$ . Six-sigma concept is used to calculate the Upper Control Limit (UCL) and Lower Control Limit (LCL) values of thresholds in order to differentiate the normal, suspicious and attack state of the total number of bytes arrival for each flow. We use the subscript 'ss' to represent 'suspicious state' and 'as' to represent 'attack state'.

### *C. Six-Sigma method used to identify threshold values:*

Six-Sigma method [24], [25] is a systematic innovative activity to statistically measure and analyze causes of defects that happen in all parts of management, and then remove those causes by identification of thresholds of the significant metrics which are measured with help of the data collected from the process. It is proposed by Motorola to address quality problem and business improvement. Six-Sigma claims that focusing on reduction of variation will solve process and business problems. By using a set of statistical tools to understand the fluctuation of a process, management can begin to predict the expected outcome of that process. If the outcome is not satisfactory, associated tools can be used to further understand the elements influencing that process. Using Six-Sigma there would be approximately 3.4 or fewer failures per billion attempts. This is an extremely low rate of failure. It has been demonstrated that six sigma methodologies, integrated with rigorous statistics, can be flexible, powerful and successful without being either overly simplistic or inordinately cumbersome [36]. To find six-sigma, calculate sigma or standard deviation, multiply by 6, and add or subtract the result to the calculated mean. Hence, to achieve extremely low false positives and negatives, six-sigma method is used in our attack flows characterization approach to identify the threshold values. Theoretical control limits of UCL and LCL for suspicious state are represented as follows:

$$UCL_{ss} = m + 3s \quad (6)$$

$$UCL_{ss} = m - 3s \quad (7)$$

where, $UCL_{ss}$ and $LCL_{ss}$ represents a 3 x sigma upwards and downwards deviation from the mean value of per flow total number of bytes arrival, respectively. For normally distributed output, 99.7% should fall between $UCL_{ss}$ and $LCL_{ss}$. Theoretical control limits of UCL and LCL for attack state are represented as:

$$UCL_{as} = m + 6s \quad (8)$$

$$UCL_{as} = m - 6s \quad (9)$$

where, $UCL_{as}$ and $LCL_{as}$ represents a 6 x sigma upwards and downwards deviation from the mean value of per flow total number of bytes arrival, respectively. For normally distributed output, 99.97% should fall between $UCL_{as}$ and $LCL_{as}$. Here $m$ and $s$ represents mean and standard deviation of per flow total bytes arrival, respectively, when system is attack free.

Therefore, flows which have total number of the bytes arrival during monitoring period is greater than $UCL_{as}$ or smaller than $LCL_{as}$ are considered attack flows. The values between $LCL_{ss}$ and $UCL_{ss}$ are considered to be under normal state. Flows, which have total number of the bytes arrival during monitoring period is between $UCL_{as}$ and $UCL_{ss}$ or between $LCL_{as}$ and $LCL_{ss}$ are considered suspicious flows. There can still be false positives and negatives due to flash crowd. To further reduce false positive negatives, flows that are active in previous time window are omitted from list of attack flows. Thus, all the packets coming through flows $F^{*}_{attack}$ are filtered at edge routers. Rate throttling strategy is applied to packets coming through flows $F^{*}_{suspicious}$. Rate of packets coming through flows $F^{*}_{suspicious}$ is throttled according to strength of attack. If incoming rate of attack traffic is high, packets coming through flows $F^{*}_{suspicious}$ are throttle with high rate and vice versa.

## IV. Experimental Setup and Performance Analysis

In this section, we evaluate our proposed scheme using simulations. The simulations are carried out using NS-2 [27] network simulator. We show that false positives and false negatives triggered by our algorithm are very less. This implies that profiles built are reasonably stable and characterizes malicious traffic correctly.

### *A. Simulation Environment*

Presently, the Internet can be viewed as a collection of interconnected routing domains, which are groups of nodes under a common administration that share routing information. A primary characteristic of these domains is routing locality, in which the path between any two nodes in a domain remains entirely within the domain. Thus, each routing domain in the Internet can be classified as either a stub or a transit domain [37], [38]. A domain is a stub domain if the path connecting nodes N1 and N2 passes through that domain and if either N1 or N2 is located in that domain. Transit domains do not have this restriction. The purpose of transit domains is to interconnect stub domains efficiently. Thus, real-world Internet type topologies generated using Transit-Stub model of GT-ITM [26] topology generator is used to test our proposed scheme, where transit domains are treated as different Internet Service Provider (ISP) network i.e. Autonomous System (AS). For simulations, we use ISP level topology, which contains four transit domains with each domain contain twelve transit nodes i.e. transit routers. All the four transit domains have two peer links at transit nodes with adjacent transit domains. Remaining ten transit nodes are connected to ten stub domain, one stub domain per transit node. Stub domains are used to connect transit domains with customer domains, as each stub domain contains a customer domain with ten legitimate client machines. So total of four hundred legitimate client machines are used to generate background traffic. Total zombie machines range between 10 and 100 to generate attack traffic. Transit domain four contains the server machine to be attacked by zombie machines. A short scale simulation topology is shown in fig. 4.

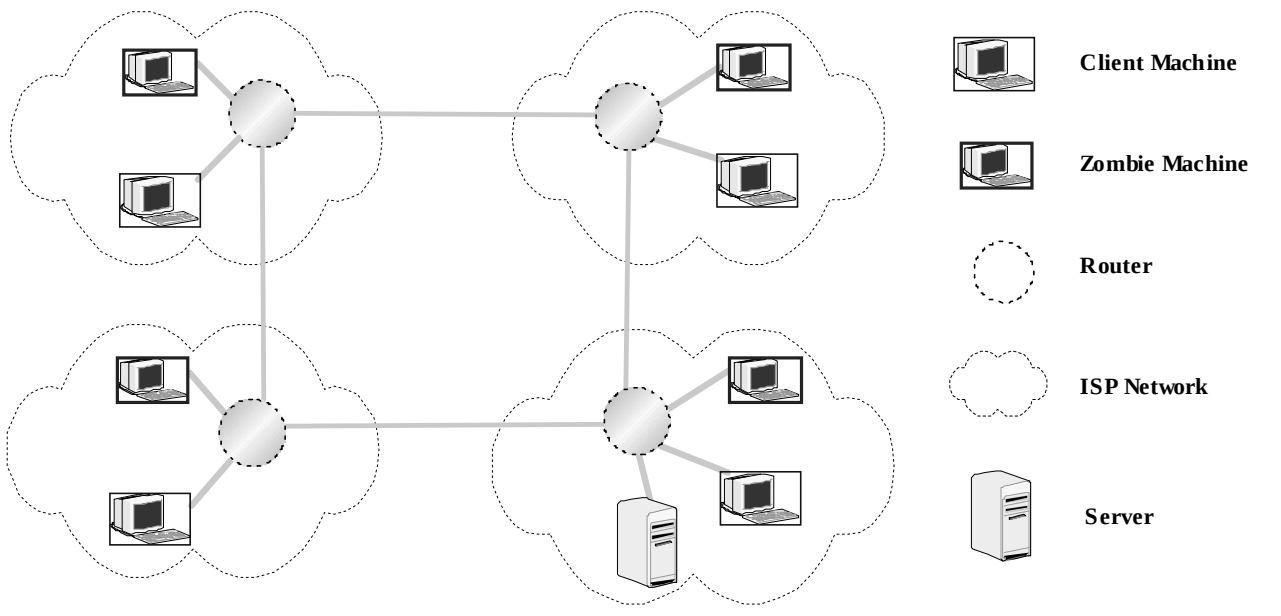

Fig. 4 A short scale simulation topology

Currently, the majority of the DDoS attacks are TCP flooding, so we will consider detection of a wide range of TCP flooding attacks in this section. The legitimate clients are TCP agents that request files of size 1 Mbps with request inter-arrival times drawn from a Poisson distribution. The attackers are modeled by UDP agents. A UDP connection is used instead of a TCP one because in a practical attack flow, the attacker would normally never follow the basic rules of TCP, i.e. waiting for ACK packets before the next window of outstanding packets can be sent, etc. The attack traffic rate varies from 0.1 to 3.5 Mbps per attack daemon. False positive alarm number increases steadily with increasing monitoring window size as shown in fig. 5. Even false positive rate is minimum using window size 100 ms but detection rate is very less i.e. 74 % using this value. Therefore, in our experiments, the monitoring time window was set to 200 ms, as the typical domestic Internet RTT is around 100 ms and the average global Internet RTT is 140 ms [39]. Total false positive alarms are minimum with high detection rate using this value of monitoring window. The simulations are repeated and different attack scenarios are compared by varying total number of zombie machines and at different attack strengths.

Fig. 6 shows temporal variation of volume metric when (a) system is in normal condition, (b) under low rate DDoS attack and (c) under high rate DDoS attack.

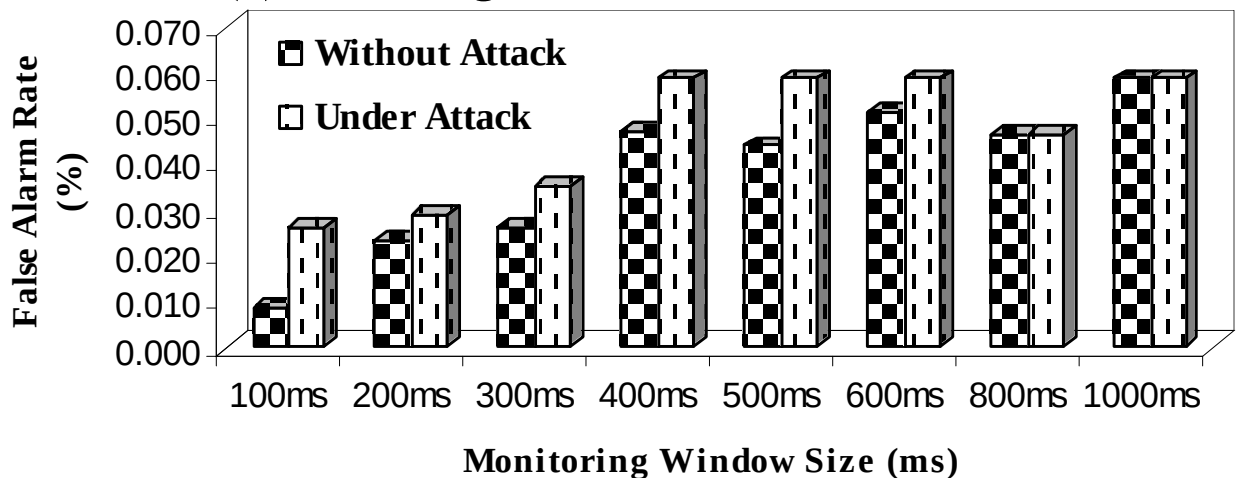

Fig. 5 Variation of false alarm rate using various window sizes

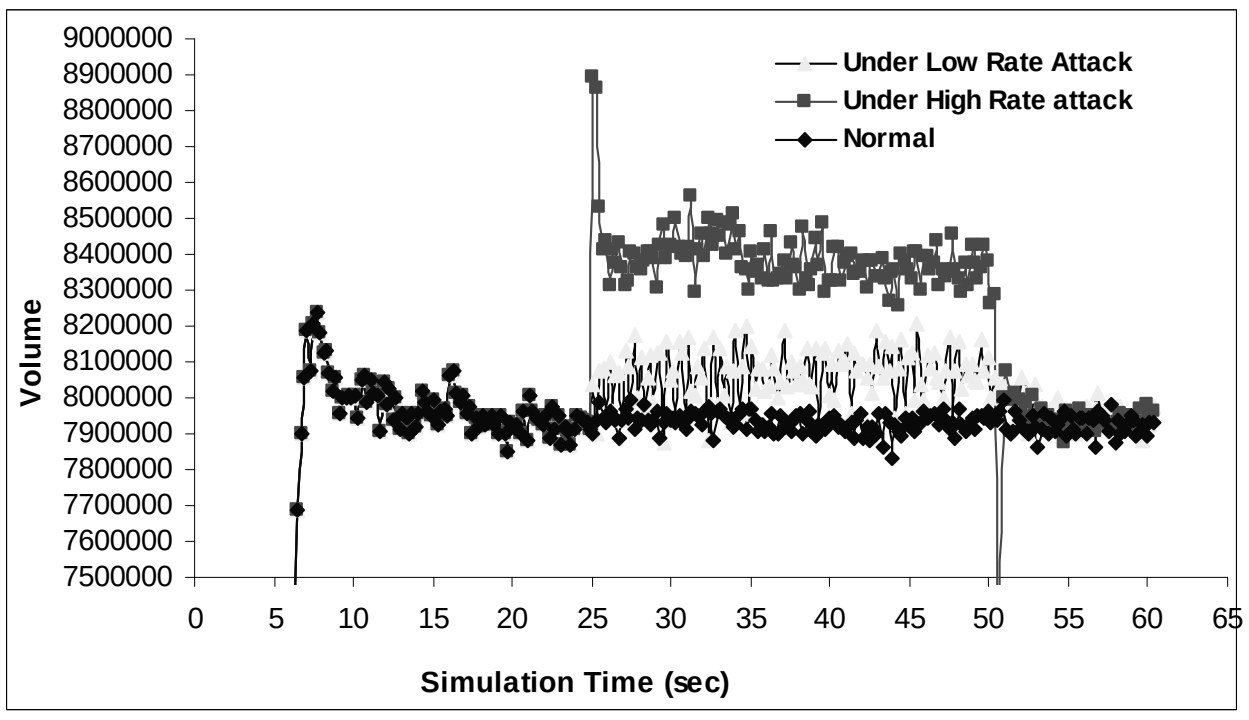

Fig. 6 Temporal variation of volume metric when system is in normal condition, under low rate DDoS attack, and under high rate DDoS attack

DDoS attacks start at 25th second and end at 50th second. 400 client machines are used to send TCP traffic. High rate attack is performed using 100 zombie machines with mean rate 3Mbps per attacker. To perform low rate attack 100 zombie machines are used with mean rate 0.1Mbps per attacker. As shown in figure, it is clear that low rate attacks are nearly undetectable when using only volume as statistical measure.

For detection of low rate DDoS attack correctly with low false positive rate, flow metric is also considered along with volume metric. Fig. 7 shows temporal variation of flow metric when (a) system is in normal condition, (b) under DDoS attack using 25, 50, 75 and 100 zombie machines. It is clear from the fig. 6 and fig. 7, that low rate DDoS attacks perform using large number of zombie machines are also easily detected using both flow and volume metrics simultaneously.

### B. *Performance Evaluation Metrics*

We have used three metrics to evaluate performance of our proposed DDoS detection approach, namely, detection rate ($R_d$), false positive alarm rate ($R_{fp}$), and receiver operating characteristic (ROC). The detection rate ($R_d$) is the measure of percentage of attacks detected among all actual attacks performed. The detection rate ($R_d$) is defined as follows:

$$R_d = d/n \quad (10)$$

where d is the number of DDoS detected attacks, and n is the total number of actual attacks generated during the simulation. The false positive alarm rate ($R_{fp}$) is the measure of percentage of false positives among all normal traffic event defined as follows:

$$R_{fp} = f/m \quad (11)$$

where f is the number of false positive alarm raised by attack detection mechanism, and n is the total number of normal traffic flow events during the simulation. The ROC curve is used to evaluate tradeoff between detection rate and false positive rate.

### C. *Simulation Results and Discussion*

As discussed earlier, effectiveness of an anomaly based detection system highly depends on accuracy of threshold value settings. Inaccurate threshold values cause a large number of false positives and false negatives. We use tolerance factors to set threshold values accurately. Tolerance factors are tunable parameters, which can tune according to network condition. Thus, it is possible that values of tolerance factors for a particular network environment are not suitable for other network. Therefore, various simulations are performed using different value of tolerance factors. Then, trade-off between detection and false positive rate provides guidelines for selecting values of tolerance factors for a particular simulation environment.

Fig. 8 illustrates the variation of the detection and false positive rate with respect to different value of detection tolerance factors $r_1$, $r_2$. We can see from the figure, as values of tolerance factors increases detection rate which was nearly 100 tend to decrease after $r_1$>=6 and $r_1$ >=6. However, false positive rate is very high for $r_1$>=5 and $r_1$ >=5. So by careful investigations, we select optimal value of $r_1$=6 and $r_2$=6, on which detection rate is close to 99% with less than 3% false positive rate. Above result demonstrates that detection rate is very high with low false positive rate when $r_1$=6 and $r_2$=6. The ROC curve in fig. 9 also shows same results.

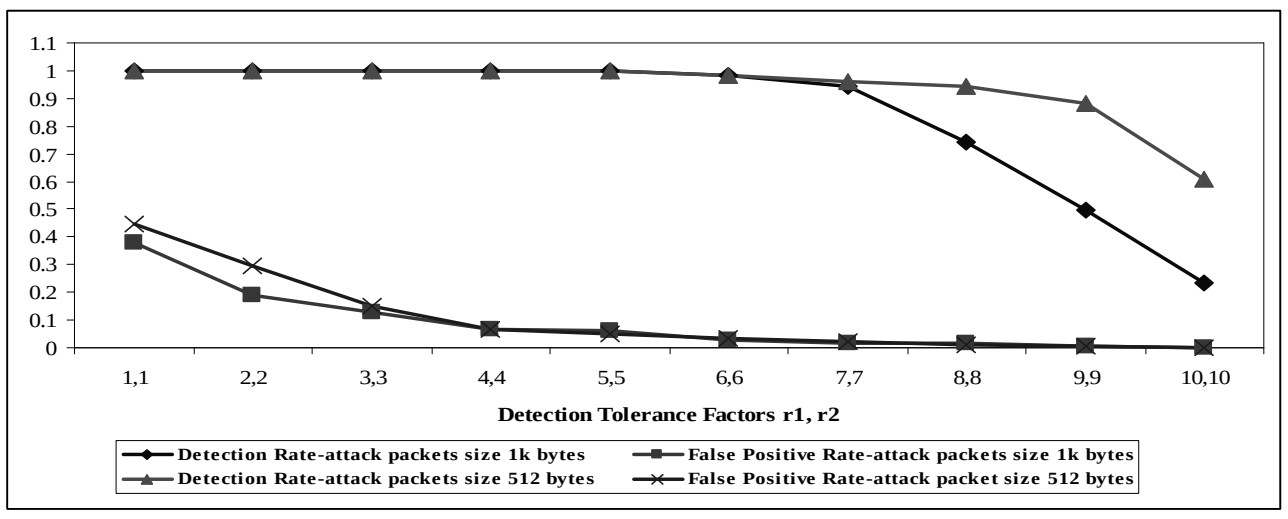

Fig. 7 Effect of detection tolerance factor on the detection and false positive rate

Therefore, value of tolerance factor $r_1$, $r_2$ is taken 6 in our approach. Values of $r_1$, $r_2$ can vary for different network conditions and correct value can be selected by drawing tradeoff between detection and false positive rate.

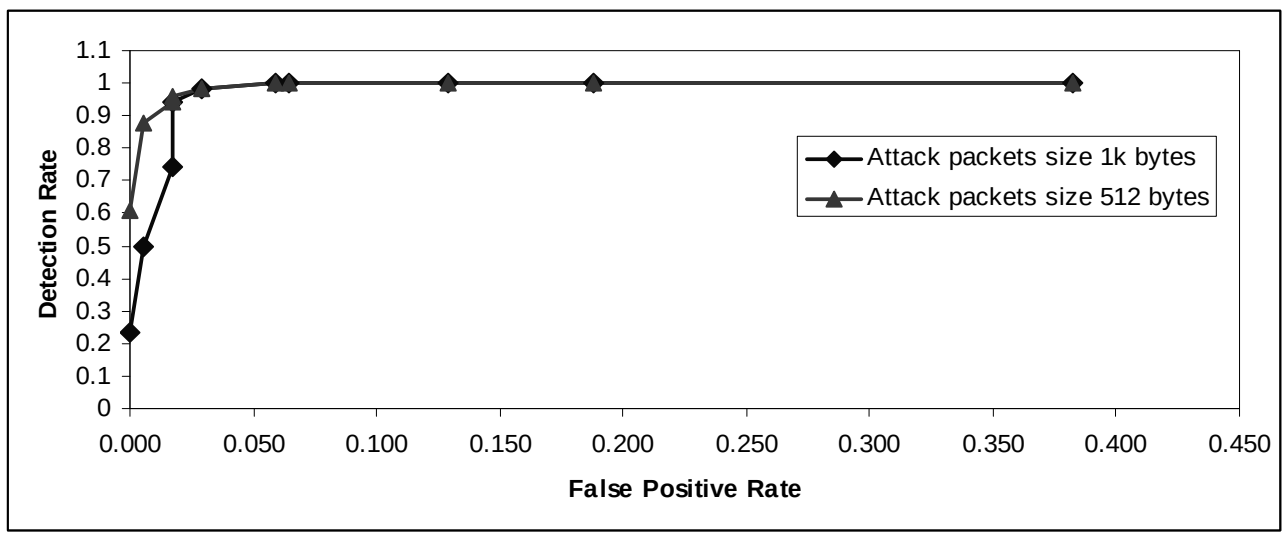

Fig. 8 ROC curve showing the tradeoff between the detection rate and false positive rate of DDoS attacks

## V. Validation with Real Data

In this section, performance of proposed FVBA scheme is evaluated on KDD 99 dataset [28], which is publicly available benchmark dataset. We first describe dataset in details, then how it preprocessed according to our specific purposes and assumptions. Then FVBA is under training and several results are displayed. Finally, we evaluate our approach with test data and discuss the results.

### *A. KDD 99 dataset description and preprocessing*

MIT Lincoln Lab's DARPA intrusion detection evaluation datasets have been employed to design and test intrusion detection systems [40]. In 1999, Stofo et. al. summarized recorded network traffic from the DARPA 98 Lincoln Lab dataset into network connections with 41-features per connection [28], [41]. This formed the KDD 99 intrusion detection benchmark dataset that is most popular dataset used to test and evaluate a large number of IDSs. KDD dataset covers following four major categories of attacks:

- Denial-of-Service (DoS) attacks (deny legitimate requests to a system), e.g. ping-of-death, SYN flood
- Probing attacks (information gathering attacks), e.g. Port scanning
- Remote-to-Local (R2L) attacks (unauthorized local access from a remote machine), e.g. guessing password
- User-to-Root (U2R) attacks (unauthorized access to local super-user or root), e.g. various buffer overflow attacks

In the present work, our focus is the detection of a wide range of DoS attacks. KDD dataset is divided into labeled and unlabeled records. Each labeled record consisted of 41 features and one target value. KDD dataset contains several data files, in which we choose two files: kddcup.data_10_percent.gz and corrected.gz. In kddcup.data_10_percent.gz, there are around 5 million (494021) records and it was used for training and validating FVBA DDoS detection system. In corrected.gz, there are around 3 million (311029) records and it was used for testing FVBA detection system.

Three types of connections are there in KDD dataset: TCP connections, UDP connections and ICMP connections. Distribution of these connections in both training and testing datasets is shown in fig. 10.

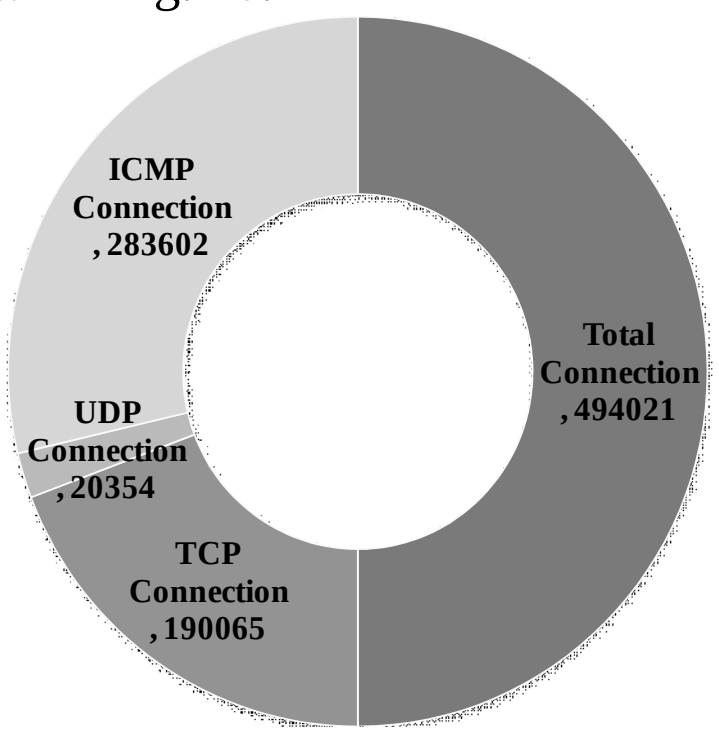

**(a)**

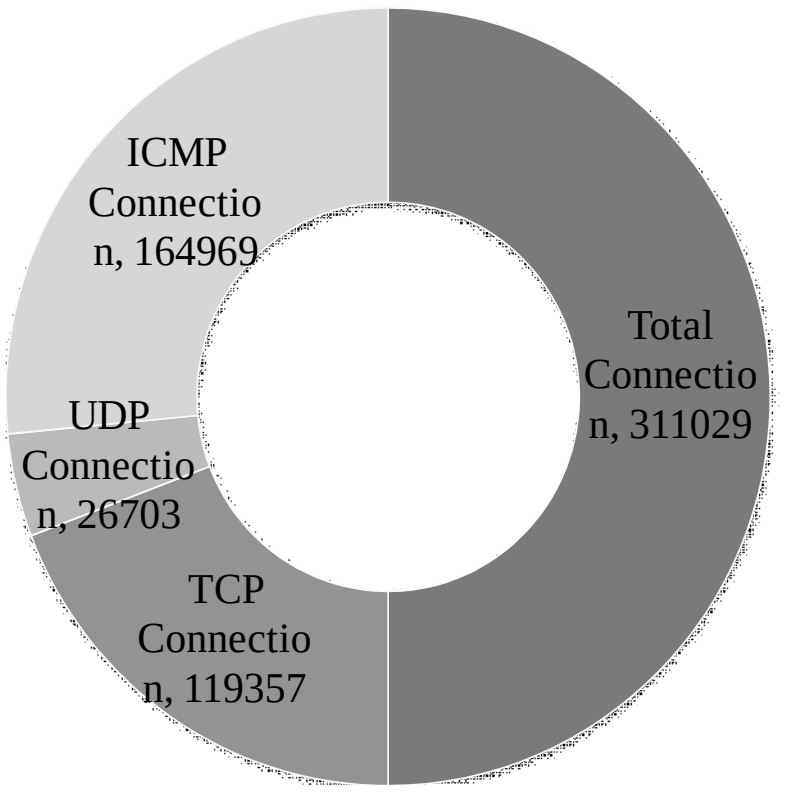

**(b)**

Fig. 9 Distribution of TCP, UDP and ICMP connections in (a) training dataset, (b) testing dataset

Total 22 and 37 attack types are there in training and testing datasets respectively. Table I list the DoS attack types, protocol categories and instances in both training and testing datasets. We can see that there are 6 and 10 different types of DoS attacks in training and testing dataset respectively.

Each record in dataset has 41 extracted features, in which 38 features are continuous and others are symbolic. There are four categories of derived features, which are 9 intrinsic features, 13 content features, 9 traffic features and 10 host features.

TABLE I: ATTACKS DISTRIBUTION IN (A) TRAINING DATASET, (B) TESTING DATASET

(a)

| DDoS Attack Types | Protocol Category | Instances |
|---|---|---|
| back | TCP | 2203 |
| land | TCP | 21 |
| neptune | TCP | 107201 |
| pod | ICMP | 264 |
| smurf | ICMP | 280790 |
| teardrop | UDP | 979 |

(b)

| DDoS Attack Types | Protocol Category | Instances |
|---|---|---|
| apache2 | TCP | 794 |
| back | TCP | 1098 |
| land | TCP | 9 |
| mailbomb | TCP | 5000 |
| neptune | TCP | 58001 |
| pod | ICMP | 87 |
| processtable | TCP | 759 |
| smurf | ICMP | 164091 |
| teardrop | UDP | 12 |
| udpstorm | UDP | 2 |

First, we filter out connection records of DoS attacks category and then remove labels from both the training and testing dataset. Then normal profile is set for each protocol category as each flow is determined by protocol. To set normal profile, volume and flow measures are calculated using training dataset.

### *B. Training*

Effectiveness of proposed detection system highly depends on accuracy of threshold value settings. Inaccurate threshold values cause a large number of false positives and false negatives. Therefore, various simulations are performed using different values of tolerance factors $r_1$ and $r_2$ and $r_3$.

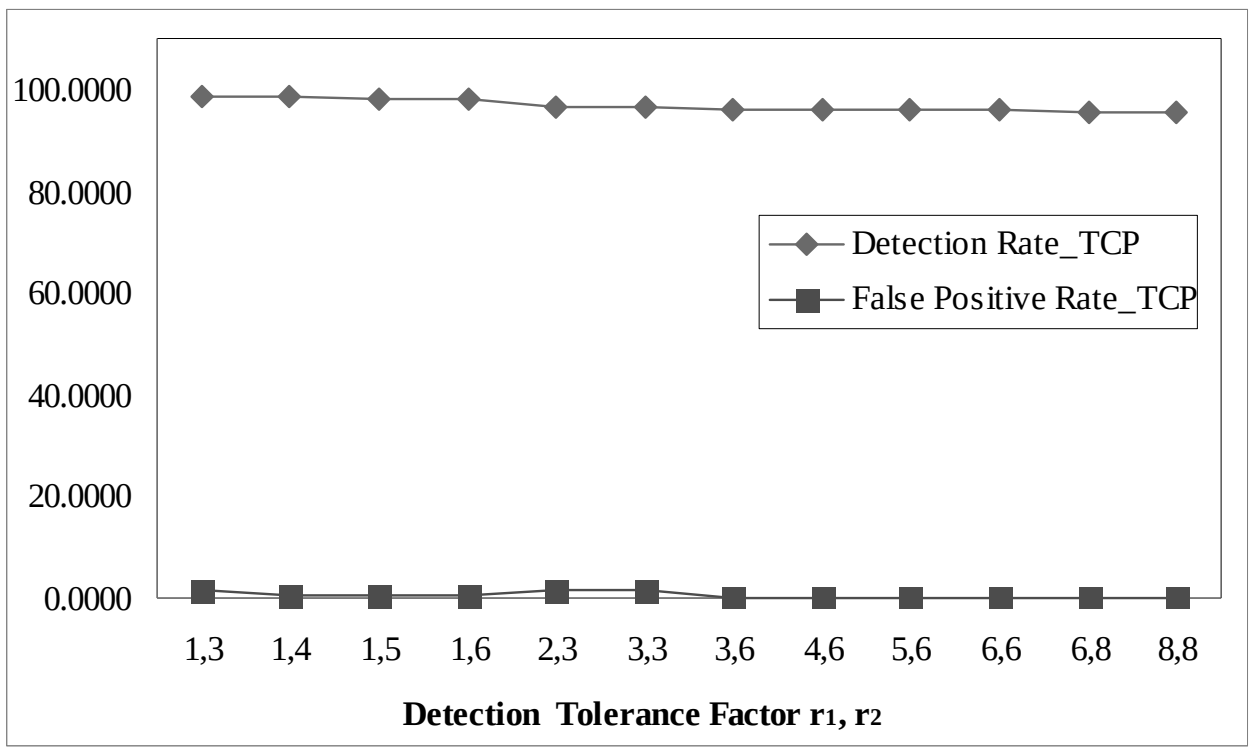

Fig. 10 Percentage of detection and false positive rates with varying tolerance factors r1 and r2 for TCP connections

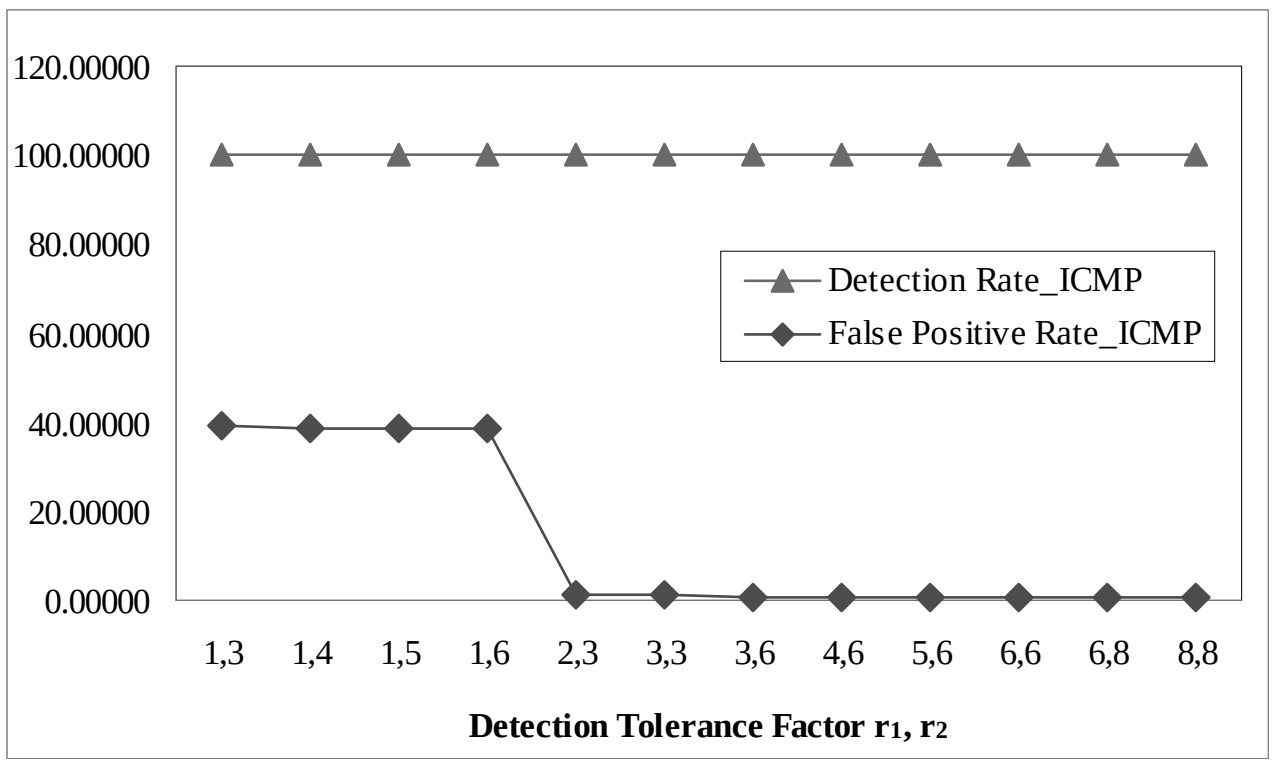

Fig. 11 Percentage of detection and false positive rates with varying tolerance factors r1 and r2 for ICMP connections

Fig. 11 and 12 illustrates the variation of the detection and false positive rate with respect to different values of detection tolerance factors $r_1$ and $r_2$ for TCP and ICMP protocol category.

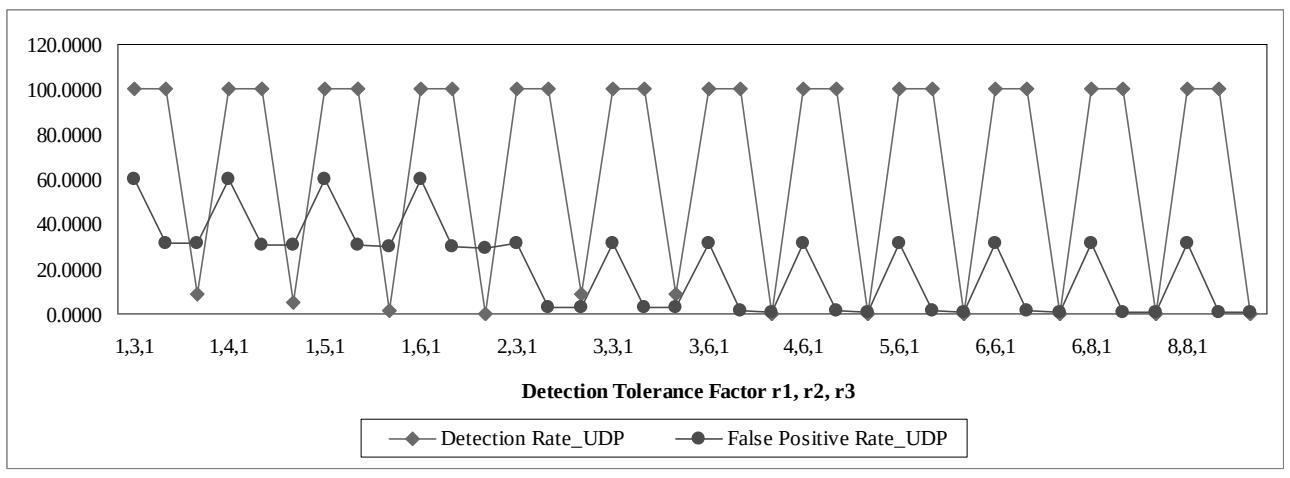

Fig. 12 Percentage of detection and false positive rates with varying tolerance factors r1, r2 and r3 for UDP connections

We can see, detection rate is 98.0827% with 0.34630% of false positives, when $r_1$ =1 and $r_2$=5 for TCP connection. For ICMP connection detection rate is 100% with 0.7764% of false positives, when $r_1$ =5 and $r_2$=6. Fig. 13 illustrates the variation of the detection and false positive rate with respect to different values of detection tolerance factors $r_1$, $r_2$ and $r_3$ for UDP protocol category.

TABLE II: OPTIMAL VALUES OF TOLERANCE FACTORS TO SET NORMAL PROFILE

| Protocol Category | Tolerance Factors value |
|---|---|
| TCP | $r_1$=1, $r_2$=5 |
| UDP | $r_1$=6, $r_2$=8, $r_3$=1.5 |
| ICMP | $r_1$=5, $r_2$=6 |

Detection rate is 100% with 0.8656% of false positives, when $r_1$ =6, $r_2$=8 and $r_3$=1.5. Therefore, optimal values of tolerance factors to set the normal profile are as given in table II.

### *C. Testing*

After training and validating, we apply proposed detection system on test dataset. Table III summarizes the overall results of testing for different protocol category.

TABLE III: OVERALL RESULTS OF TESTING

| Protocol Category | Total Detected | Detection Rate (%) |
|---|---|---|
| TCP Connection | 58675/65661 | 89.36 |
| UDP Connection | 14/14 | 100 |
| ICMP Connection | 164178/164178 | 100 |
| overall | 222867/229853 | 96.9 |

Table IV contains summary of different types of DoS attacks detected in test dataset. Above stated results show that our proposed framework yields 96.9 percent detection accuracy with less than 1 percent false alarms.

TABLE IV: DOS ATTACKS DETECTION SUMMERY IN TEST DATASET

| Attack Types | Protocol Category | Total Detected | Detection Rate (%) |
|---|---|---|---|
| apache2 | TCP | 634/794 | 79.84 |
| back | TCP | 1068/1098 | 97.26 |
| land | TCP | 0/9 | 0 |
| mailbomb | TCP | 0/5000 | 0 |
| neptune | TCP | 56973/58001 | 98.22 |
| pod | ICMP | 87/87 | 100 |
| processtable | TCP | 0/759 | 0 |
| smurf | ICMP | 164091/164091 | 100 |
| teardrop | UDP | 12/12 | 100 |
| udpstorm | UDP | 2/2 | 100 |

## VI. CONCLUSION

In this paper, a novel framework is proposed that autonomously detects and accurately characterizes a wide range of DDoS attacks, ensuring good service to legitimate clients. Consideration of varying tolerance factors make proposed detection system scalable to the varying network conditions and attack loads in real time. Six- sigma method enables accurate characterization of malicious flows from attack flows.

In addition to controlled test-bed experiments, effectiveness of the proposed system is verified through intensive experiments with KDD 99 dataset. Proposed system has demonstrated an excellent performance in both test-bed experiments and in the real operation. It is found that combining flow and volume measures are better way to find signs of attack as compared to volume or entropy measure.

We have implemented our approach in single ISP network but it can easily be deployed at multiple ISPs with help of trusted entities acting as interfaces between two ISPs so that two ISPs can share there information and thus more effectively stop the attack. The enormous complexity of DDoS problem requires a comprehensive solution that encompasses multiple stages of the process of defense against DDoS attacks. Therefore, investigation of an accurate response strategy to complete the framework and strengthen the defense against DDoS attacks is a future research issue.

## ACKNOWLEDGMENT

The authors gratefully acknowledge the financial support of the Ministry of Human Resource Development (MHRD), Government of India for partial work reported in the paper.